\documentclass[aps,prl,twocolumn,showpacs,psfig,superscriptaddress,longbibliography]{revtex4-1}

\usepackage{textcomp}
\usepackage{times}
\usepackage{graphicx}
\usepackage{float}
\usepackage{latexsym,amsmath,amssymb,bm,euscript}
\usepackage{color}
\usepackage{subfigure}
\usepackage{epstopdf}
\usepackage[colorlinks=true,linkcolor=blue,citecolor=blue]{hyperref}
\usepackage{hyperref}
\usepackage{soul}
\usepackage[normalem]{ulem}
\usepackage{mathrsfs}
\usepackage{amsmath}
\usepackage{lettrine}
\usepackage{xspace}
\usepackage{textcomp}

\newcommand {\B}{\textcolor {blue}}

\newcommand{\Fig}[1]{Fig.~\ref{#1}}

\def\para{\ensuremath{/\kern -0.8em /}\xspace}

\graphicspath{{../Figs/}}

\begin{document}

\title{Evidence of the Berezinskii-Kosterlitz-Thouless Phase in a Frustrated Magnet}

\author{Ze~Hu}
\thanks{These authors contributed equally to this study.}
\affiliation{Department of Physics and Beijing Key Laboratory of
Opto-electronic Functional Materials $\&$ Micro-nano Devices, Renmin
University of China, Beijing, 100872, China}

\author{Zhen~Ma}
\thanks{These authors contributed equally to this study.}
\affiliation{National Laboratory of Solid State Microstructures and Department
of Physics, Nanjing University, Nanjing 210093, China}
\affiliation{Institute for Advanced Materials, Hubei Normal University, Huangshi 435002, China}

\author{Yuan-Da~Liao}
\thanks{These authors contributed equally to this study.}
\affiliation{Beijing National Laboratory for Condensed Matter Physics, and Institute of Physics, Chinese Academy of Sciences, Beijing 100190, China}
\affiliation{School of Physical Sciences, University of Chinese Academy of Sciences, Beijing 100190, China}

\author{Han~Li}
\thanks{These authors contributed equally to this study.}
\affiliation{School of Physics, Key Laboratory of Micro-Nano Measurement-Manipulation and
Physics (Ministry of Education), Beihang University, Beijing 100191, China}

\author{Chunsheng~Ma}
\affiliation{Department of Physics and Beijing Key Laboratory of
Opto-electronic Functional Materials $\&$ Micro-nano Devices, Renmin
University of China, Beijing, 100872, China}

\author{Yi~Cui}
\affiliation{Department of Physics and Beijing Key Laboratory of
Opto-electronic Functional Materials $\&$ Micro-nano Devices, Renmin
University of China, Beijing, 100872, China}

\author{Yanyan~Shangguan}
\affiliation{National Laboratory of Solid State Microstructures and Department
of Physics, Nanjing University, Nanjing 210093, China}

\author{Zhentao~Huang}
\affiliation{National Laboratory of Solid State Microstructures and Department
of Physics, Nanjing University, Nanjing 210093, China}

\author{Yang~Qi}
\email{qiyang@fudan.edu.cn}
\affiliation{Center for Field Theory and Particle Physics, Department of Physics, Fudan University, Shanghai 200433, China}
\affiliation{State Key Laboratory of Surface Physics, Fudan University, Shanghai 200433, China}
\affiliation{Collaborative Innovation Center of Advanced Microstructures, Nanjing University, Nanjing 210093, China}

\author{Wei~Li}
\email{w.li@buaa.edu.cn}
\affiliation{School of Physics, Key Laboratory of Micro-Nano Measurement-Manipulation and Physics (Ministry of Education),
 Beihang University, Beijing 100191, China}
\affiliation{International Research Institute of Multidisciplinary Science, Beihang University, Beijing 100191, China}

\author{Zi~Yang~Meng}
\email{zymeng@hku.hk}
\affiliation{Department of Physics and HKU-UCAS Joint Institute of Theoretical and Computational Physics, The University of Hong Kong, Pokfulam Road, Hong Kong SAR, China}
\affiliation{Beijing National Laboratory for Condensed Matter Physics, and Institute of Physics, Chinese Academy of Sciences, Beijing 100190, China}
\affiliation{Songshan Lake Materials Laboratory, Dongguan, Guangdong 523808, China}

\author{Jinsheng~Wen}
\email{jwen@nju.edu.cn}
\affiliation{National Laboratory of Solid State Microstructures and Department
of Physics, Nanjing University, Nanjing 210093, China}
\affiliation{Collaborative Innovation Center of Advanced Microstructures, Nanjing University, Nanjing 210093, China}

\author{Weiqiang~Yu}
\email{wqyu\_phy@ruc.edu.cn}
\affiliation{Department of Physics and Beijing Key Laboratory of
Opto-electronic Functional Materials $\&$ Micro-nano Devices, Renmin
University of China, Beijing, 100872, China}

\begin{abstract}
\end{abstract}

\date{\today}
\maketitle

\noindent {\bf{Abstract}}\\
\textbf{
The Berezinskii-Kosterlitz-Thouless (BKT) mechanism,
building upon proliferation of topological defects in 2D systems,
is the first example of phase transition beyond the Landau-Ginzburg paradigm
of symmetry breaking. Such a topological phase transition has long been
sought yet undiscovered directly in magnetic materials.
Here, we pin down two transitions that bound a BKT phase in an ideal 2D
frustrated magnet TmMgGaO$_4$,
via nuclear magnetic resonance under in-plane magnetic fields,
which do not disturb the low-energy electronic states and allow
BKT fluctuations to be detected sensitively.
Moreover, by applying out-of-plane fields, we find a critical scaling
behaviour of the magnetic susceptibility expected
for the BKT transition. The experimental findings can be explained
by quantum Monte Carlo simulations applied
on an accurate triangular-lattice Ising model of the compound
which hosts a BKT phase. These results provide
a concrete example for the BKT phase
and offer an ideal platform for future investigations
on the BKT physics in magnetic materials.
}

\noindent {\bf{Introduction}}\\
Topology
plays an increasingly important role in understanding different phases and phase transitions
in correlated quantum matters and materials.
One prominent example is the BKT mechanism in 2D systems
\cite{Berezinskii1971,Berezinskii1972,Kosterlitz1972,Kosterlitz1973,Kosterlitz1974},
which is associated with the binding and unbinding of topological defects.
The BKT transition cannot be characterized by conventional order parameters,
and constitutes the earliest example of phase transition beyond the Landau-Ginzburg
paradigm of spontaneous symmetry breaking.
Historically, the BKT mechanism was introduced in the $XY$ spin model and long predicted
to occur in magnetic thin films~\cite{Berezinskii1971,Kosterlitz1973}.
In experiments, signatures of the BKT transition have been observed in superfluid helium films~\cite{Bishop1978},
as well as 2D superconductor films~\cite{Hebard1980,Epstein1981} and arrays~\cite{Resnick1981}.
Regarding the original proposal in layered $XY$-type magnets,
despite intensive efforts~\cite{Heinrich2003,Cuccoli2003,Wawrzy2008,Wheeler2009,Tutsch2014,Kumar2019},
direct and unambiguous observation of the BKT transition is still lacking.
One major obstacle is the 3D couplings in the magnets,
although weak, will inevitably enhance the confining potential of vortices~\cite{Kumar2019}, leading to 3D ordering that masks the BKT transition.
Therefore, it is of fundamental interest to find and identify BKT materials that
could overcome the obstacle and study the topology-related low-energy dynamics.

\begin{figure}[t]\includegraphics[width=8.5cm]{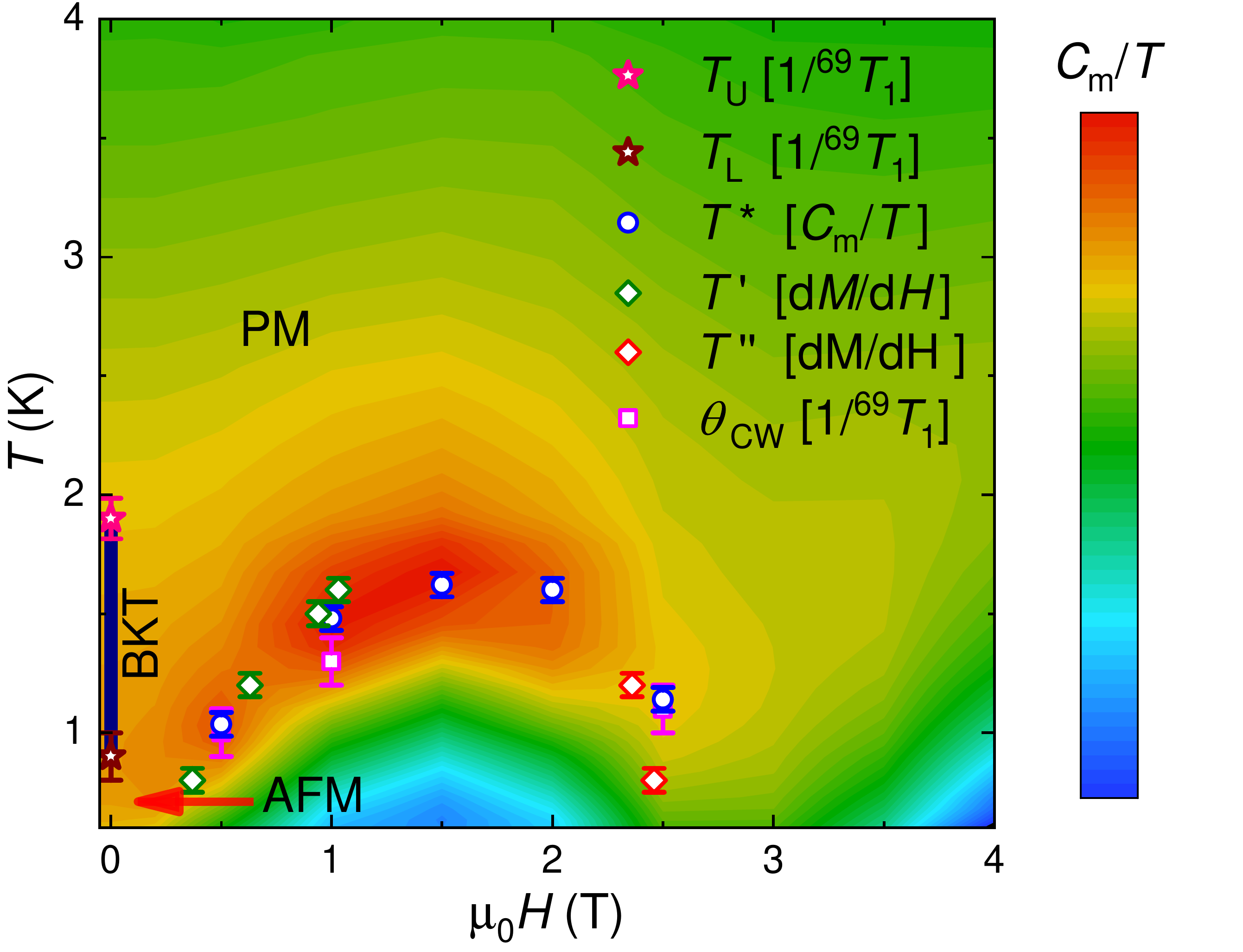}
\caption{{\bf Phase diagram of TmMgGaO$_4$ under out-of-plane magnetic fields.} Under zero field, there are paramagnetic (PM), BKT, and antiferromagnetic (AFM) phases. The $T_{\rm U}$ ($T_{\rm L}$) is the upper (lower) BKT transition temperature, determined from the plateau structure in the NMR spin-lattice relaxation rate $1/^{69}T_1$ (see Fig.~\ref{fig:fig2}{\bf c} for details). The BKT phase between $T_{\rm U}$ and $T_{\rm L}$ is illustrated by the solid vertical line, while the AFM regime is indicated by the arrow. The contour background depicts the magnetic specific heat $C_{\rm m}/T$ at various fields and temperatures, with data adapted from Ref.~\cite{Li2020} and plotted in logarithmic scale. $T^*$ corresponds to the maximum of $C_{\rm m}/T$ at each field, signifying the position of strong magnetic fluctuations. $T'$($T''$) denotes the temperature at a specific field where a peak is found in the differential susceptibility $dM/dH$, shown in Fig.~\ref{fig:fig3} {\bf b}. The Curie-Weiss temperature $\theta_{\rm CW}$ is obtained from the $1/^{69}T_1T$ (see Supplementary Fig.~1). Remarkably, $T^*$, $T'$, $T''$ and $\theta_{\rm CW}$ all collapse to the same phase boundary between the BKT-like regime and AFM phase. A magnetically ordered phase is supposed to lie below the dome-like boundary. Errors represent one standard deviation throughout the paper.}\label{fig:fig1}
\end{figure}

\begin{figure*}[t]
	\includegraphics[width=12cm]{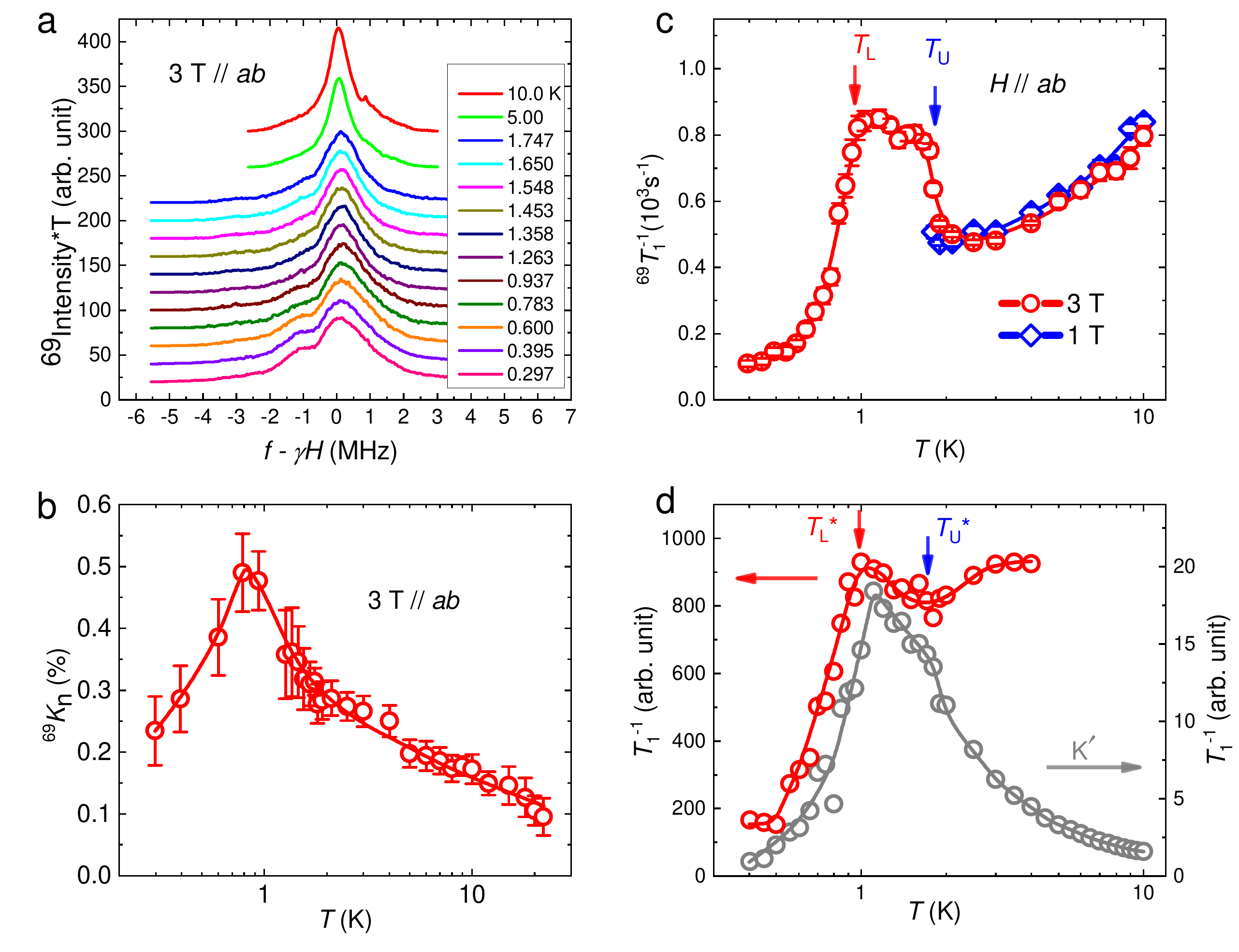}
	\caption{\label{fig:fig2}
	{\bf NMR spectra and spin-lattice relaxation rates of TmMgGaO$_4$.}
	{\bf a} $^{69}$Ga NMR spectra at different
	temperatures under a 3~T in-plane field,
	with zero frequency corresponding to 
	${\gamma}H$=~30.692 MHz.
        Data are shifted vertically for clarity.	
        At high temperatures, the spectra are roughly symmetric,
        while for $T \leq $ 2~K, a shoulder-like structure
        can be resolved on the left side of the main peak.
	{\bf b} NMR hyperfine shift 
	$^{69}K_n = ({\bar{f}}/{\gamma H} - 1) \times 100\%$
	as a function of temperature,
	where $\bar{f}$ is the average frequency of each spectrum.
	{\bf c} NMR spin-lattice relaxation rate $1/^{69}T_1$
	versus temperature measured under in-plane fields of 3~T and 1~T.
	A plateau-like feature, characterizing strong magnetic fluctuations,
	is observed between $T_{\rm L} \simeq 0.9$~K (lower BKT transition)
	and $T_{\rm U} \simeq1.9$~K (upper BKT transition).
	{\bf d} $1/T_1$ data computed from the dynamical
	spin-spin correlation function with contributions from 
	all momentum points 
	[c.f., Eq.~(\ref{Eq:Ahf}), and see the hyperfine form factor  
	in Supplementary Note~\B{2}]
	in the Brillouin zone (left scale)
	and from $\rm K'$ point in the vicinity of the K point (right scale),
	through large-scale
	QMC simulations (see Methods).
	}
\end{figure*}

Recently, a layered frustrated rare-earth antiferromagnet TmMgGaO$_4$~\cite{Cava2018,Shen2019,Li2020}
was reported to ideally realize the triangular-lattice quantum Ising (TLI) model~\cite{Li2020a}.
The relatively large interlayer distance of 8.3774~\r{A} along the $c$ axis
gives rise to excellent two dimensionality~\cite{Cava2018}, and no sign of conventional
3D phase transition was observed in either specific heat or magnetic susceptibility measurements.
Nevertheless, it was reported from neutron scattering that
TmMgGaO$_4$ ordered below $\sim$1~K into an antiferromagnetic phase
with a six-fold symmetry breaking~\cite{Shen2019,Li2020}, which closely resembles the ground state
of the TLI model originated from an order-by-disorder mechanism~\cite{R.Moessner2001,Isakov2003}.
At higher temperatures, the effective $XY$ degrees of freedom emerge,
and the BKT mechanism is expected to come into play~\cite{Isakov2003}.

In TmMgGaO$_4$, each Tm$^{3+}$, with a $4f^{12}$
electron configuration and a spin-orbit moment $J=6$,
forms a non-Kramers doublet due to the crystal-electric-field splitting.
The doublet is well separated from the rest levels by about 400~K~\cite{Li2020},
and can thus be regarded as an effective spin-1/2.
There is further a fine energy splitting within the doublet,
induced by the local trigonal crystal field~\cite{Cava2018},
acting as an intrinsic ``transverse field" applied on the effective spin.
From the magnetization measurements~\cite{Cava2018,Shen2019,Li2020},
one observes that Tm$^{3+}$ ions contribute highly anisotropic
Ising-type moments with $J_z=\pm 6$ along the $c$ axis,
resulting in an effective out-of-plane $g$-factor  $\sim$13.2~\cite{Li2020,Li2020a}.
On the contrary, the effective in-plane $g$-factor and
dipolar moment in the $ab$ plane are negligible.

Facilitated by this feature in TmMgGaO$_4$, in this work
we employed nuclear magnetic resonance (NMR),
a sensitive low-energy probe, to detect the BKT phase.
We applied a moderate in-plane field of 3~T,
which is adequate to collect the $^{69}$Ga NMR signals and,
at the same time, hardly disturbs the low-energy electronic states of the material.
This is important, since in the TLI model that is believed
to accurately model TmMgGaO$_4$~\cite{Li2020a},
the BKT phase can be fragile against out-of-plane fields
~\cite{Liu2019arXiv,Damle2015,Biswas2018},
thus posing a challenge to NMR measurements.
Taking advantage of the fact that in-plane moment in TmMgGaO$_4$
is mostly multipolar~\cite{Li2020,Li2020a},
our NMR experiments with in-plane fields, which merely couples to the nuclear spins,
can clearly identify the BKT phase in the material.

As shown in Fig.~\ref{fig:fig1},
from our NMR measurements of the spin-lattice relaxation rate $1/T_1$,
we identify $T_{\rm U} \simeq 1.9$~K and $T_{\rm L} \simeq 0.9$~K
which represent the upper- and lower-BKT transitions,
where a critical BKT phase
resides at zero magnetic field in between the high-$T$
paramagnetic and low-$T$ antiferromagnetic phases.
This finding is further substantiated by our scaling analysis
of the measured susceptibility data near $T_{\rm L}$,
as well as the simulated NMR and susceptibility data using large-scale
quantum Monte Carlo (QMC) calculations.

\bigskip \noindent {\bf Results}\\
\textbf{NMR probe of the BKT phase.}
The obtained NMR spectra with an in-plane magnetic field $\mu_0 H = 3$~T
are shown in Fig.~\ref{fig:fig2}{\bf a} at representative temperatures.
In order to better resolve the magnetic transition,
the hyperfine shifts $^{69}K_{\rm n}$ of the spectra were analyzed
and plotted in Fig.~\ref{fig:fig2}{\bf b} as a function of temperature.
Upon cooling, $^{69}K_{\rm n}$ peaks at about 0.8-0.95~K,
and then starts to drop at lower temperatures. Therefore, the
ordering temperature is determined to be $T_{\rm L}~{\simeq}~0.9$~K,
consistent with neutron scattering experiments~\cite{Shen2019,Li2020}.
Additionally, both the second moments (width of the NMR spectra)
and the third moments (asymmetry of the spectra)
of the spectra change dramatically below $\sim$~2~K,
suggesting the onset of local hyperfine fields
enhanced by the static or quasi-static magnetic ordering
(Supplementary Fig.~4). These two characteristic temperatures signal
the two-step melting of magnetic order
through two BKT transitions, suggesting an
intermediate floating BKT phase in the system. We suspect that there is some inhomogeneity of the local hyperfine fields, which
is very likely caused by the quenched disorder from
Mg/Ga site mixing~\cite{Li2020}, although
no significant influence on the electronic
and more importantly the magnetic
properties is seen (see more detailed discussions in
Supplementary Note~\B{2}).

The spin-lattice relaxation rate $1/T_1$ provides a highly sensitive detection
of low-energy spin fluctuations~\cite{Julien2000,Kitagawa2008,
Koutroulakis2015,Jeong2017,Moriya2003},
and thus the BKT transition. In Fig.~\ref{fig:fig2}{\bf c} we show the $1/^{69}T_1$
obtained under in-plane fields $\mu_0H = 3$~T  and $1$~T,
which reflects intrinsic spin fluctuations with zero out-of-plane field.
At 3~T, $1/^{69}T_1$ first decreases upon cooling from 10~K
then suddenly increases below $T_{\rm U}\approx$ 1.9~K, indicating
the onset of strong low-energy spin fluctuations. The data at 1~T show similar behaviours.
Below $T_{\rm L}\simeq0.9$~K, $1/^{69}T_1$ drops sharply, consistent with the onset
of the magnetic ordering as also inferred from the hyperfine shift.
Here, $1/^{69}T_1$ is dominated by the gapped
long wavelength excitations in the order state.
At the magnetic phase transition, a peaked feature
in $1/T_1$ develops, caused by the gapless low-energy
spin fluctuations with diverging correlation length.
Remarkably, at temperatures between
$T_{\rm U}\simeq1.9$~K and $T_{\rm L}\simeq0.9$~K, $1/^{69}T_1$ exhibits a
plateau-like structure, indicating the emergence of
a highly fluctuating phase with diverging spin correlations yet no true long-range order,
which is the hallmark of a BKT phase
~\cite{Berezinskii1971,Berezinskii1972,Kosterlitz1972,Kosterlitz1973,Kosterlitz1974}.
Therefore, it is for the first time that such
a phase is unambiguously observed in a magnetic crystalline material.

Our unbiased QMC simulations on the TLI model of the material (see Methods),
with accurate coupling parameters determined in Ref.~\cite{Li2020a},
quantitatively justifies the existence of the
BKT phase between $T_{\rm L}$ and $T_{\rm U}$.
We computed $1/T_1$ and compared with the experiment below.
Figure~\ref{fig:fig2}{\bf d} shows the calculated $1/T_1$ by including
fluctuations from all momentum points in the Brillouin zone
(cf. Supplementary Fig.~2), and compare to that from $\rm K'$
(around the K point at the corner of Brillouin zone).
The former shows a decreases upon cooling below 4~K,
and then an upturn above
$T_{\rm U}^* \simeq 2$~K, followed by a rapid decrease
below $T_{\rm L}^* \simeq 1$~K.
These behaviours are in excellent agreement with the
measured $1/^{69}T_1$.
The latter reflects gapless excitations of
the $XY$ degrees of freedom emergent in the BKT phase,
where the calculated $1/T_1$ from $\rm K'$ exhibits
an anomalous increase down to $T_{\rm U}^*$,
below which the increment slows down.
The contribution to $1/T_1$ near the K point reaches a
maximum at $T_{\rm L}^*$ and drops rapidly below it.
The absence of critical spin fluctuations at momentum
away from the K point suggests that the measured
$1/^{69}T_1$ below 2~K is mainly contributed by
excitations around the K point (see Supplementary Note~\B{3}).

Overall, the quasi-plateau behaviours in the QMC results,
as well as the two characteristic temperature scales,
are in full consistency with the NMR measurements.
This constitutes both strong support for the accurate
quantum many-body modeling of the material TmMgGaO$_4$
and also solid proof of the BKT phase therein detected by NMR.
Nevertheless, we note that there are still subtle 
differences between the experimental and numerical data. 
Needless to say, the real material is always more complicated 
than our theoretical minimal model.
For example, influences from higher CEF levels
above the non-Kramers doublet,
the interlayer couplings not included in our model calculations,
and the lack of knowledge on the precise local 
hyperfine coupling constant, etc.,
may explain the difference remained between the
panels (c) and (d) of Fig.~\ref{fig:fig2}.

\begin{figure*}[t]
	\includegraphics[width=18cm]{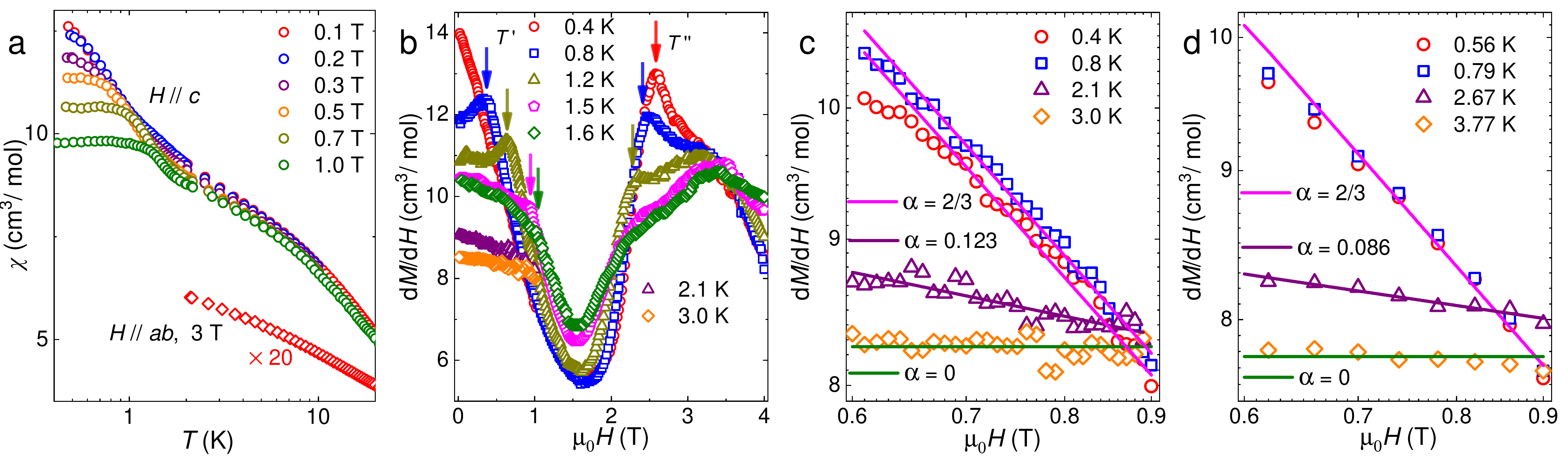}
	\caption{\label{fig:fig3}
	{\bf Uniform magnetic susceptibility of TmMgGaO$_4$ and scaling analysis.}
	{\bf a} dc susceptibility $\chi$ as functions of temperatures
		under small out-of-plane ($H~\para~c$) and in-plane ($H~\para~ab$) fields.
		The latter is multiplied by a factor of 20 for visualizing purpose.
		The deviation of data below 2~K indicates the entry to
		the BKT phase and the field-suppression of magnetic correlations.
		{\bf b} The differential susceptibility ${dM}/{dH}$
		under out-of-plane fields at different temperatures.
		The kinks at low fields, as denoted by the down arrows,
		suggest the transition from the BKT-like phase to the
		ordered phase (under the dome
		in Fig.~\ref{fig:fig1}) with increasing fields.
		The peaked features at high fields suggests a
		quantum phase transition to the polarized phase.
		{\bf c} Fits of ${dM}/{dH}$ to the power-law scaling
		function $dM/dH\sim H^{-\alpha}$ with
		$\alpha=2/3$ for the 0.4 and 0.8-K data,
		and $\alpha=0.123$ for 2.1~K data in the log-log scale.
		The 3~K data follow the $\alpha=0$ line in the
		paramagnetic phase.
		{\bf d}  ${dM}/{dH}$ by the QMC calculations in the same
		field and temperature range as in {\bf c}, and
                 fits to the power-law function with exponents $\alpha$,
		which give consistent results as experiments.
	}
\end{figure*}


\noindent\textbf{Universal magnetic susceptibility scaling.}
Magnetic susceptibility $\chi$ measurements are also performed to
strengthen the finding of the BKT phase.
In Fig.~\ref{fig:fig3}{\bf a}, we show the overall
temperature dependence of $\chi$
at different out-of-plane fields. For $T\gtrsim2$~K, $\chi$ increases monotonically
upon cooling, and barely changes with fields. However, for $T\lesssim 2$~K,
approximately the upper BKT transition $T_{\rm U}$ as obtained
from the $1/^{69}T_1$ measurements, $\chi$ increases as the field decreases,
suggesting the onset of peculiar magnetic correlations.
With further cooling, the susceptibility gets flattened with temperature.
The magnetization $M(H)$ is further measured at selected temperatures
(data shown in Supplementary Fig.~6),
and for the sake of clarity, the differential susceptibility ${dM}/{dH}$ is plotted
in Fig.~\ref{fig:fig3}{\bf b}. At around 2.5~T, a pronounced peak can be
observed at low temperature, indicating the existence
of a quantum phase transition, and the phase at lower fields
should be a magnetically order phase,
although its precise nature remains to be uncovered.
Besides the high-field feature, for temperatures at 0.8~K and above,
a kinked feature is clearly resolved on each ${dM}/{dH}$ curve at low fields;
whereas at 0.4~K, the low-field kink disappears,
which posts a question of whether there
is a quantum transition or merely a crossover from the zero-field
AFM phase to the finite-field ordered phase
under the dome in Fig.~\ref{fig:fig1}.
The temperature and field values indicated by the
down arrows in Fig.~\ref{fig:fig3} are denoted as
$T'$ and $T''$ in the phase diagram (Fig.~\ref{fig:fig1}).

Field-theoretical analysis of the TLI model~\cite{Damle2015,Biswas2018}
has predicted that upon applying a small out-of-plane field,
the differential susceptibility $dM/dH$ shall exhibit
a divergent power-law behaviour as ${dM}/{dH}\sim H^{-\alpha}$
in proximity to the BKT phase.
At $T_{\rm L}$, $\alpha$ has the value of ${2}/{3}$,
which corresponds to a critical exponent $\eta=1/9$ at the lower BKT transition
and is originated from the six-fold symmetry breaking~\cite{Damle2015}.
The exponent $\alpha$ gradually decreases as temperature increases,
and above an intermediate temperature between $T_{\rm L}$ and $T_{\rm U}$,
$\alpha = 0$ due to non-universal contributions.
This is exactly what we observe in Fig.~\ref{fig:fig3}{\bf c}.
We fit the $dM/dH$ with the power-law function at different temperatures,
with the fitting regime chosen between 0.6-0.9~T. At 0.8~K,
$\alpha$ is very close to the expected value of $2/3$ (and thus $\eta=1/9$) at $T_{\rm L}$,
which constitutes a remarkable fingerprint evidence for the BKT transition.
At lower temperatures such as 0.4~K, the exponent is also close to $2/3$ because the
susceptibility saturates with temperature, as shown in Fig.~\ref{fig:fig3}{\bf a}.
At high temperatures, $\alpha$ drops rapidly to a small value 0.12 at 2.1~K,
and becomes effectively zero at 3.0~K.

Therefore, the
susceptibility scaling gives the lower BKT transition at about 0.8~K and
upper transition probably between 2.1 and 3~K, in good agreement
with the $T_{\rm L}$ and $T_{\rm U}$ estimated from NMR.
These results are also fully consistent with
our QMC calculations on the susceptibility shown in
Fig.~\ref{fig:fig3}{\bf d}. At $T_{\rm L}$ or lower,
$\alpha$ is ${2}/{3}$, then decreases to a very small exponent $0.086$
at 2.67~K, and above 3~K, becomes zero within numerical uncertainty.
Such a power-law behaviour in ${dM}/{dH}$ again signifies the finite-temperature
window of the BKT phase with diverging magnetic correlations,
which gives rise to the universal power-law scaling of magnetic susceptibility.


\bigskip \noindent {\bf Discussion}\\
We believe the findings in this work are of various fundamental values.
Since the original proposal of a BKT phase in
magnetic films~\cite{Kosterlitz1972,Kosterlitz1973,Kosterlitz1974},
which also triggered the currently thriving
research field of topology in quantum materials,
tremendous efforts have been devoted to finding the
BKT phase in magnetic crystalline materials,
yet hindered by the obstacle outlined in the Introduction.
Here, benefiting from NMR as a sensitive low-energy probe,
and the nearly zero planar gyromagnetic factor
in a TLI magnet TmMgGaO$_4$, we are able to
reveal two BKT transitions and a critical BKT phase
with an emergent $XY$ symmetry.
Together with the power-law behaviour of the susceptibility,
and excellent agreement between
our QMC simulation and experiment data,
we unambiguously identify the long-sought BKT
phase in a magnetic crystalline material.

Many intriguing questions are stimulated,
based on the phase diagram in Fig.~\ref{fig:fig1}
obtained here. First, for example, what is the nature
of the ordered phase under finite fields, are there
further exotic phases and transitions in the phase diagram,
and will there be a field-induced quantum phase transition
at the high-field side of the dome -- these are all of great interests
to be addressed in future studies.
Second, it should be noted that the dynamical properties
obtained by QMC calculations in \Fig{fig:fig2} are computed on
a large, while finite-size, 36$\times$36 lattice, which already produces
$1/T_1$ data in very good agreement with the experimental measurements.
Such a great agreement is surprising,
given the possible existence of randomness
from Ma/Ga site mixing in the material TmMgGaO$_4$ \cite{Li2020},
and also the lattice disorder revealed by the large high-temperature
second moment of the NMR spectra (Supplementary note~\B{4}).
Although the random distribution in intrinsic transverse
fields and spin couplings does not seem to
alter the low-temperature spin ordered
phase as well as the sharp spin excitation line shapes
\cite{Shen2019,Li2020a}, its intriguing
effects on the finite-temperature phase diagram of TLI
and also the compound TmMgGaO$_4$ is calling for further studies.

Third, in the study of BKT transition in superfluid systems,
it has been observed experimentally and understood theoretically
that additional dissipations also appear above the
transition temperature due to fluctuations of vortices~\cite{Bishop1978}.
Hence, the plateau of $1/T_1$ we observe may also cover
regions slightly above the upper BKT transition temperature.
We leave this subtlety to future numerical and experimental efforts.
Lastly, in general terms, whether there are other rare-earth magnetic
materials in the same family of TmMgGaO$_4$ that,
acquire similar 2D competing magnetic interactions
from highly anisotropic gyromagnetic factor and unique triangular-lattice structures,
and also exhibit the BKT physics, is quite intriguing and calls for future investigations.
All these directions are ready to be explored from here.\\

\bigskip \noindent {\bf Methods}\\
\textbf{Crystal growth and susceptibility measurements.}
Single crystals were grown by the optical-floating-zone 
method with an image furnace
(IR Image Furnace G2, Quantum Design).
The natural cleavage surface of the crystals is the $ab$ plane,
which allows us to align the field orientation
within an error of $2^{\circ}$. The dc susceptibility was measured
in a PPMS VSM (Quantum Design) for temperatures above
2~K and in a He-3 MPMS (Quantum Design) for temperatures ranging
from 0.4 to 2~K.  \\

\noindent 
\textbf{NMR measurements.} 
The $^{69}$Ga ($I$~=~${3}/{2}$, $\gamma$~=~10.219~MHz/T) 
NMR spectra were collected with the standard spin-echo sequence,
with frequency sweep by a 50 kHz step using a topping tuning circuit.
The NMR hyperfine shift was obtained by calculating
the change of the first moment of the spectra.
The spin-lattice relaxation rate $1/^{69}T_1$
was measured by the inversion-recovery method, where a $\pi/2$ pulse
was used as the inversion pulse. The NMR data from 1.8~K and above were measured
in a variable temperature insert, and the data from 1.8~K and below were measured in
a dilution refrigerator. The weak NMR signal at low fields and the rapid decrease of
$^{69}T_2$ upon cooling (Supplementary Note~\B{5}) prevented us to measure the $1/^{69}T_1$
for in-plane fields less than 3~T, with temperature below 1.8~K.
Whereas for in-plane fields of 4~T and higher, the sample could not be held
in position because of the large anisotropy in the $g$ factor and unavoidable
sample misalignment ($\lesssim2^\circ$).   At $T=$1.2~K, we did not find any
change of $1/^{69}T_1$ with two different radio frequency excitation levels (14~mT and 24~mT),
and with different frequencies across the NMR line, within the error bar. \\

\noindent 
\textbf{Triangular lattice Ising model.}
At zero field, the intralayer couplings
in TmMgGaO$_4$ can be described by the TLI Hamiltonian,
\begin{eqnarray}
\hspace{-9mm}&&H = J_1\sum_{
\langle i,j \rangle}S_i^zS_j^z + J_2\sum_{\langle\langle i,j\rangle\rangle}S_i^zS_j^z + \sum_i\Delta S_i^x,
 \label{h1}
\end{eqnarray}
where $J_1$ and $J_2$ are the superexchange interactions among Tm$^{3+}$,
$\langle i,j \rangle$ and $\langle\langle i,j \rangle\rangle$ refer to the
nearest neighbors and the next-nearest neighbors, respectively,
and $\Delta$ is the energy splitting within the non-Kramers doublet imposed by the crystal field.
We have shown that the parameter set $J_1$$=$ 0.99~meV,
$J_2/J_1$$\approx$~0.05 and $\Delta/J_1\approx$0.54
reproduces the experimental results of the
transition temperatures and the
inelastic neutron scattering spectra~\cite{Li2020a}.

In the TLI model [Eq.~(\ref{h1})],
we can define a complex field $\psi$ as a
combination of the Ising ($Z$) components
$m_{\rm A,B,C}^z$ on three sublattices, i.e.,
\begin{equation}
\psi = m^z_{\rm A} + {\rm e}^{i2\pi/3} \, m^z_{\rm B}  +  {\rm e}^{i 4\pi/3} \, m_{\rm C}^z.
\label{Eq:psi}
\end{equation}
Notably, $\psi = |\psi| {\rm e}^{i \theta}$ is a complex order parameter
that represents the emergent $XY$ degree of freedom
relevant to the BKT physics in the TLI model.\\

\noindent 
\textbf{Quantum Monte Carlo calculations.}
QMC simulations were performed in the
path integral in the $S^{z}_{i,\tau}$
basis with discretization in space and time.
The lattice of $L\times L\times L_{\tau}$,
where $L=36$ and $L_{\tau}=\beta/\Delta\tau$
with $\Delta\tau=0.05$ and $\beta=1/T$,
were simulated with both local and Wolff-cluster
update schemes~\cite{YCWang2017,WLJiang2019}.
The $1/T_1$ results were obtained by first
computing the dynamical spin-spin correlation function
$\langle S^{z}_{i}(\tau)S^{z}_{j}(0)\rangle$,
and then acquiring its real-frequency dependence $S(\mathbf{q},\omega)$
from the stochastic analytic continuation~\cite{Sandvik2016}.
We then determined the $1/T_1$ either by
summing the contributions close to momentum
K or over the entire Brillouin zone, as discussed in the
Fig.~\ref{fig:fig2}d of the main text,
\begin{equation}
T^{-1}_1 (\bold{q}) = \frac{1}{L^2}\sum_{\bold q} |A_{\rm hf}({\bold q})|^2
S({\bold q},\omega \to 0),
\label{Eq:Ahf}
\end{equation}
where $A_{\rm hf}({\bold q})$ is the hyperfine coupling form factor
(see Supplementary Note~\B{2}). Similar analyses have been
successfully applied to the QMC computation
of NMR ${1}/{T_1}$ for the spin-1/2 and spin-1
chains~\cite{Sandvik1995,Capponi2019}.\\

\bigskip \noindent \textbf{Data availability}\\
The data that support the findings of this study are available from the corresponding author upon reasonable request.

\bigskip \noindent \textbf{Code availability}\\
All numerical codes in this paper are available upon request to the authors.\\


%

\bigskip \noindent {\bf Acknowledgements}\\
We thank Changle Liu, Rong Yu, Nvsen Ma and Anders Sandvik for stimulating discussions. We acknowledge the supports from the National Key
Projects for Research and Development of China through Grant Nos.~2016YFA0300502 and 2016YFA0300504, the National Natural Science Foundation of China through Grant Nos.~11574359, 11674370, 11822405, 11674157, 11974036, 11834014, 11874115 and 51872328, RGC of Hong Kong SAR China through Grant No.~17303019, Natural Science Foundation of Jiangsu Province with Grant No.~BK20180006, and Fundamental Research Funds for the Central Universities with Grant No.~020414380117, and the Research Funds of Renmin University of China. We thank the Center for Quantum Simulation Sciences in the Institute of Physics, Chinese Academy of Sciences, the Computational Initiative at the Faculty of Science and the Information Technology Services at the University of Hong Kong, the Platform for Data-Driven Computational Materials Discovery at the Songshan Lake Materials Laboratory, Guangdong, China and the Tianhe-I, Tianhe-II and Tianhe-III prototype platforms at the National Supercomputer Centers in Tianjin and Guangzhou for their technical support and generous allocation of CPU time.\\

%

\newpage
\clearpage
\onecolumngrid
\mbox{}
\begin{center}
\textbf{\large{Supplementary Information for\\
Evidence of the Berezinskii-Kosterlitz-Thouless Phase in a Frustrated Magnet}\\
Hu \textit{et al}.}
\end{center}

\setcounter{equation}{0}
\setcounter{figure}{0}
\setcounter{table}{0}
\setcounter{subsection}{0}

\renewcommand{\theequation}{\arabic{equation}}
\renewcommand{\thefigure}{\arabic{figure}}
\renewcommand{\bibnumfmt}[1]{[#1]}
\renewcommand{\citenumfont}[1]{#1}

\setcounter{equation}{0}
\setcounter{figure}{0}
\setcounter{table}{0}
\setcounter{section}{0}
\setcounter{page}{1}
\setcounter{subsection}{0}

\makeatletter

\textbf{Supplementary Note 1: Determination of the $\theta_{\text{CW}}$
from the spin-lattice relaxation rate under out-of-plane fields}

The spin-lattice relaxation rate $1/T_1$ quantifies the magnetic fluctuations through
\begin{eqnarray}
\hspace{-9mm}&& \frac{1}{T_1} = T \sum\limits_{{\bf q}} |A_{\rm hf}({\bf q})|^2 \frac{\textrm{Im}\chi({\bf q},\omega)}{\omega},
 \label{h1}
\end{eqnarray}
where $A_{\rm hf}$, Im$\chi({\bf q},\omega)$, and $\omega$ are the hyperfine coupling constant, the imaginary
part of the dynamic spin susceptibility, and the NMR frequency, respectively.

The temperature dependence of $1/^{69}T_1$ were measured
under various out-of-plane fields,
and the data are shown in \Fig{fig:fig1}.
At fields $\mu_0H=0.5$~T and 1~T, $1/^{69}T_1$ shows a weak temperature dependence from 3~K to 10~K.
However, for $\mu_0H=$ 2.5~T, $1/^{69}T_1$ increases quickly above 3~K,
indicating strong field-induced fluctuations whose origin needs to be addressed in future studies.

Below 2~K, $1/^{69}T_1$ data in \Fig{fig:fig1} show an
upturn for all measured fields,
indicating the onset of low-energy spin fluctuations upon cooling.
We fit the $1/T_1T$ with the Curie-Weiss function~\cite{Moriya2003}
$1/T_1T\sim 1/(T-\theta_{\text{CW}})$ (see the inset of \Fig{fig:fig1}),
and provide an estimate of the transition temperature $\theta_{\text{CW}}$
where the relaxation rate $1/T_1$ diverges due to critical fluctuations.
We then collect the estimated $\theta_{\text{CW}}$ at various fields and
plot them in Fig.~1 of the main text, which are in excellent agreement
with the transition temperatures determined by other means.
In particular, $\theta_{\text{CW}}$ follows a dome-like shape with fields,
in accordant to the peak positions of $C_{\rm m}/T$ and $dM/dH$ (Fig.~1 of the main text).
Note the Curie-Weiss form of the $1/T_1T$ has been widely
observed in correlated materials. It was originally introduced
by T. Moriya et al.~\cite{Moriya2003} in studying quasi-2D itinerant antiferromagnets,
where $1/T_1T\propto{\chi({\bm Q})}$, with $\chi({\bm Q})\approx (T-T_{\rm N})$ being the dynamical susceptibility at the magnetic wave vector ${\bm Q}$.
\\

\textbf{Supplementary Note 2: Hyperfine coupling form factor}

As a local probe, the NMR relaxation rate $1/T_1$ sums up
the magnetic fluctuations from
all momenta [see Eq.~(\ref{h1})], weighted by the hyperfine coupling form
factor 

\begin{equation}
|A_{\rm hf}({\bf q})|^2 = |\sum_{j} \tilde{A}_{\rm hf} e^{i{\bf q}(r_j-r_i)}|^2,
\label{eq:Ahf}
\end{equation}
where $\tilde{A}_{\rm hf}$ is the local hyperfine coupling constant,
$r_{i(j)}$ labels the position of Ga(Tm) ions,
and $j$ is NN sites of site $i$ (see Supplementary 
Fig.~\ref{fig:formfactor}a ).
By assuming an isotropic, constant $\tilde{A}_{\rm hf}=1$,
we compute the form factor   
$|A_{\rm hf}(k_x, k_y)|^2  = 1+ 4\cos\frac{k_x}{2}(\cos\frac{k_x}{2}+\cos\frac{\sqrt{3}k_y}{2})$, 
and show the results in Supplementary Fig.~\ref{fig:formfactor}b.
From the results, we find that the K point has a very small
form factor (actually zero for isotropic $\tilde{A}_{\rm hf}$)
while the M point, on the contrary, corresponds to a larger form factor.
\\

\textbf{Supplementary Note 3: Contributions to $1/T_1$ from momenta other than the K point}

\begin{figure}[t]
	\includegraphics[width=8.5cm]{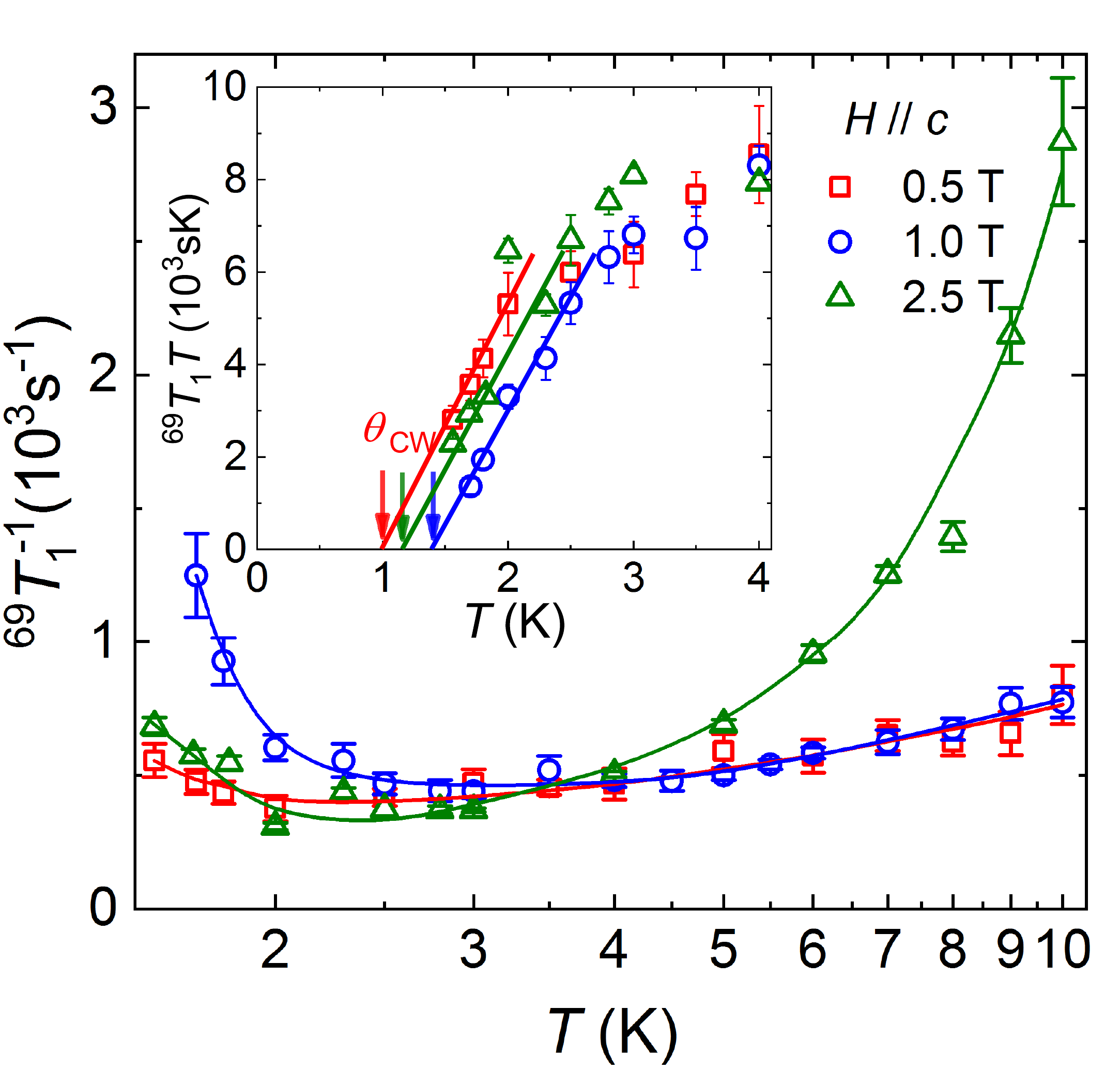}
	\renewcommand{\figurename}{\textbf{Supplementary Figure}}
	\caption{\label{fig:fig1} {\bf Spin-lattice relaxation rate $1/^{69}T_1$
		under out-of-plane fields}.
		The $1/^{69}T_1$ data are shown as functions of temperatures
		at three fields, 0.5, 1, and 2.5~T, which
		are nearly identical at temperatures between 3 and 4~K, exhibiting
		similar paramagnetic spin fluctuations. Upon cooling, an upturn appears below 3~K,
		indicating the onset of low-energy magnetic fluctuations toward the magnetic
		ordering. The inset shows the $^{69}T_1T$ plotted against temperature,
		with a Curie-Weiss fitting below 2.5~K, which provides
		an estimated temperature $\theta_{\rm CW}$
		(noted by down arrows) at which
		 spin fluctuations are maximal. These temperatures are plotted in Fig.~1 in the main text.
	}
\end{figure}

\begin{figure}[t]
\includegraphics[width=14cm]{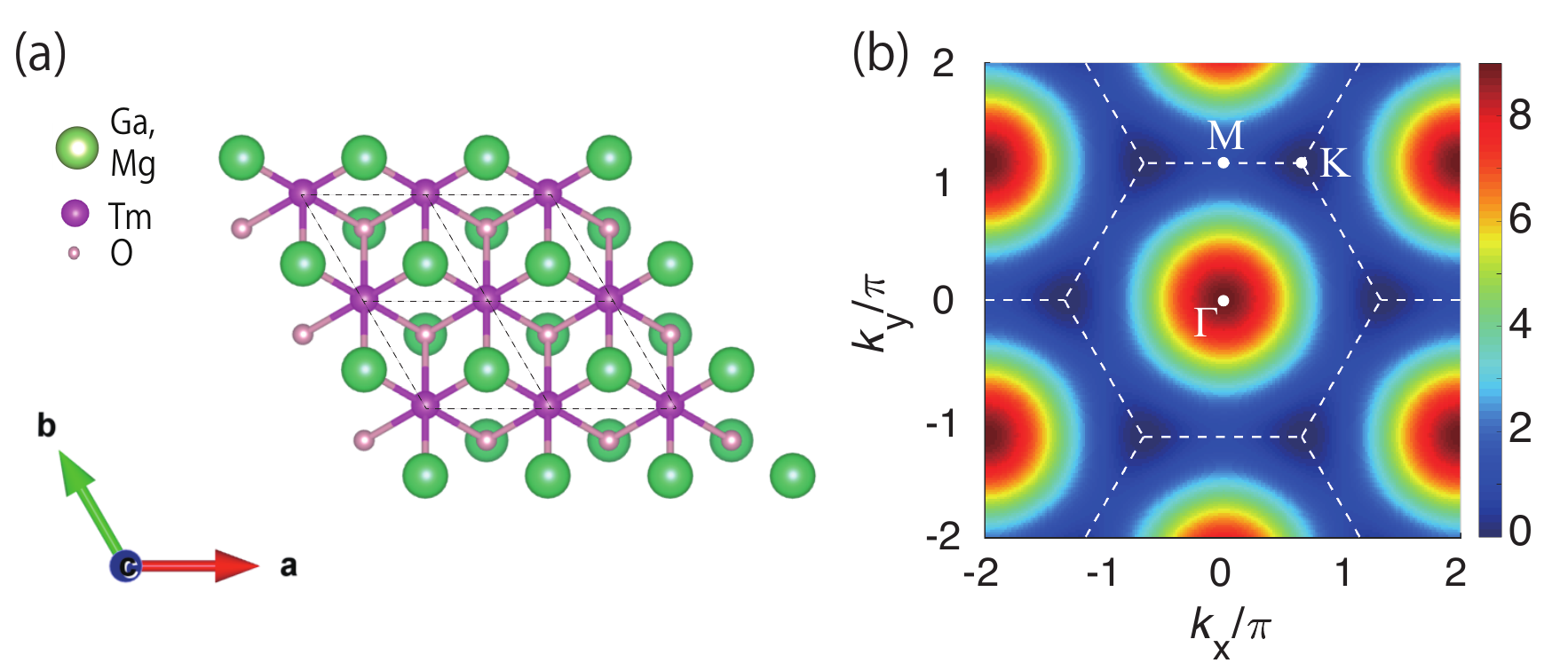}
\renewcommand{\figurename}{\textbf{Supplementary Figure}}
\caption{(a) Lattice structure within the triangular plane where
	each Ga ion has three nearest Tm sites.
	(b) The contour plot of corresponding form factor
	$|A_{\rm hf}(\bf q)|^2$ considered in the main text.}
\label{fig:formfactor}
\end{figure}

\begin{figure}[t]
\includegraphics[width=8.5cm]{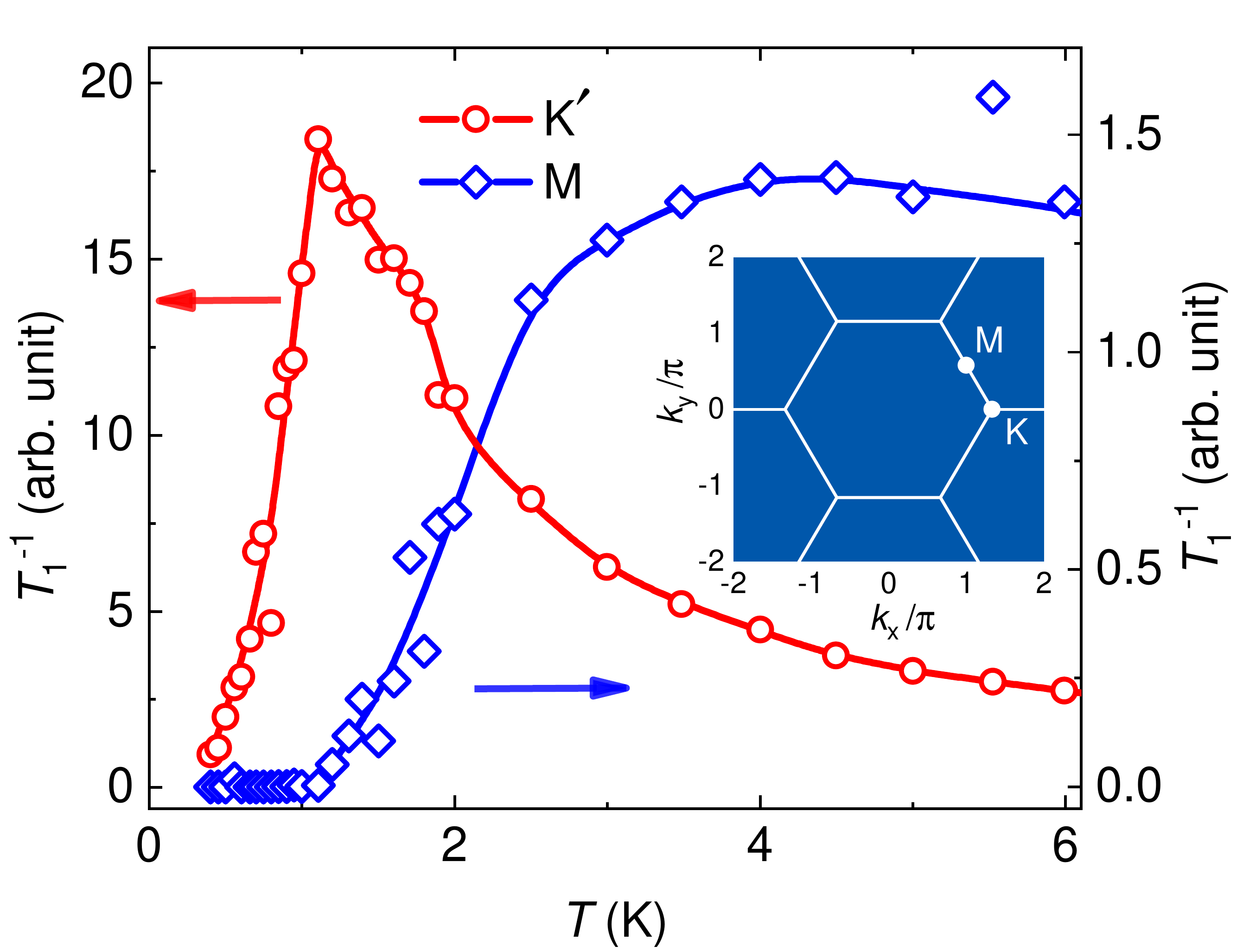}
\renewcommand{\figurename}{\textbf{Supplementary Figure}}
\caption{The spin-lattice relaxation rate $1/T_1$ calculated
by the quantum Monte Carlo simulations, including data measured
at the M point (diamonds) and a vicinity point of K (circles),
respectively. Inset illustrates
the M and K points in the Brillouin zone of the reciprocal space.
}
\label{fig:fig5}
\end{figure}

To compare with experiments,
we performed quantum Monte Carlo simulations of the $1/T_1$ (see Methods), considering
contributions from magnetic fluctuations at typical momentum points M and K
(see the inset of \Fig{fig:fig5}).
The calculated results are plotted in \Fig{fig:fig5} for comparison.
Note that M and K are the magnetic ordering wave-vectors
of the stripe and the clock antiferromagnetic phases, respectively~\cite{Li2020a}.
From the M point contribution, $1/T_1$ is
significant only at relatively high temperatures,
which drops dramatically at 2~K and is negligible when cooled down to below 1.2~K.
In contrast, the $1/T_1$ contributed from the K$'$ point is dramatically
enhanced below 2~K and peaked at about 1~K, as discussed in the main text.
The distinctive temperature dependence of the $1/T_1$ contributions
from these two momentum points
suggests a competing mechanism at low temperatures.
This can be understood because the clock phase
(fluctuations centred at K) wins over the stripe phase
(fluctuations centred at M) at $T_{\rm L}$.
From the comparisons between  numerics and experiments,
we conclude that the measured $1/^{69}T_1$ [Fig.~2(c) in the
main text] is dominated by the fluctuations with momentum
centred around K at low temperatures.
The $1/T_1$ contributions from M points, etc.,
start to play a role as temperature is escalated to above 2~K,
due to the large density of states near the M point
and the different hyperfine form factor $A_{\rm hf}({\bf q})$, etc.
\\


\begin{figure}[h]
\includegraphics[width=10cm]{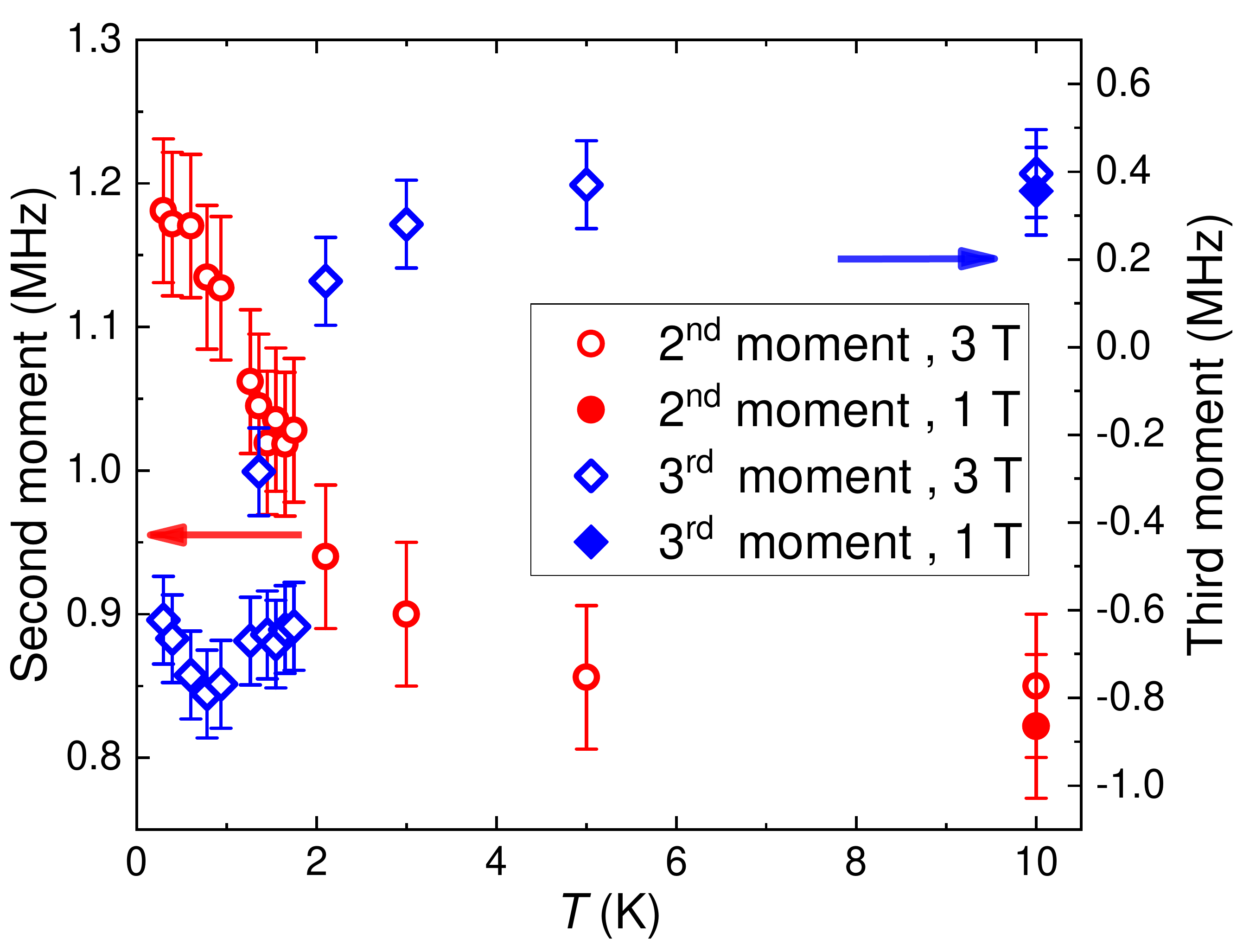}
\renewcommand{\figurename}{\textbf{Supplementary Figure}}
\caption{The second and the third moment of the
NMR spectra as functions of temperatures,
under 1~T and 3~T field in the $ab$-plane.}
\label{fig:moment}
\end{figure}

\textbf{Supplementary Note 4: Second and third moments
of the $^{69}$Ga NMR spectra}

The second and the third moments deduced from the $^{69}$Ga
NMR spectra, with 1~T and 3~T in-plane fields (i.e., $H//ab$),
are plotted in the Supplementary Fig.~\ref{fig:moment}.
The second moment, defined as $\langle (f-\overline{f})^2 \rangle^{1/2}$,
 which measures the standard deviation of the NMR spectra,
roughly equals the full-width at half-maximum of the NMR peak.
The second moment is about 0.85~MHz at $T$=10~K,
much larger than the $1/^{69}T_2$ (4~kHz)
under the same field (see Supplementary Fig.~\ref{fig:ssrr}),
indicating an inhomogeneous broadening of the spectra.
At 5~K and above, the linewidth does not change
much with temperature (cf. the data at 5~K and 10~K),
nor with field (cf. the results under 1~T and 3~T),
indicating that the high-temperature
broadening does not correspond to a magnetic origin.
Rather, since $^{69}$Ga has a 3/2 nuclear spin,
the width of the line is supposed to be defined by the unresolved
quadrupolar splitting of the line 
(into three lines for one spin-3/2 nucleus),
and therefore reflects the distribution of local 
electric-field-gradient (EFG)/quadrupolar
couplings, which in principle
barely changes with fields and temperatures.
Therefore, the high-temperature broadening of the spectra
indicates strong EFG inhomogeneity,
mostly likely due to the Mg/Ga site mixing
in the compound.

Below 2~K, on the other hand, the dramatic
broadening of the NMR linewidth upon cooling
indicates the the inhomogeneity of the local hyperfine field,
as it first occurs when the quasi-static
magnetic ordering develops below the BKT transition,
and then in the magnetically ordered phase below 0.9~K.
The frozen Mg/Ga site mixing,
ought to also enhance this inhomogeneity,
since the hyperfine field and/or the dipolar field produced by the Tm$^{3+}$ magnetic ions on the
$^{69}$Ga nucleus is affected by its exact position
and the neighbouring ions.
However, it remains an open question to which extent the local
moment of the $^{69}$Ga, and consequently the
magnetic properties, are affected by the site mixing.
This is in particular thought-provoking, as
the thermodynamic properties
of TmMgGaO$_4$ can be well fitted by the TLI model
without randomness, the magnetic order phase
is maintained at low temperature, and the
spin-wave dispersion remains sharp as observed
in INS measurements.

The third moment, defined as $\langle (f-\overline{f})^3 \rangle^{1/3}$,
measures the asymmetry of the NMR spectra.
When cooled below 2~K, the third moment
undergoes a sign change from positive to negative 
(Supplementary Fig.~4),
which actually reflects the shoulder-like feature 
developed to the left of the main peak of the spectra 
[Fig.~2(a) in the main text].
We ascribe this asymmetry of spectra to the dipolar 
hyperfine field with static/quasi-static ordering.
However, to find out the accurate origin of these 
features and nail down the corresponding
magnetic structures is beyond the scope of present 
NMR measurements.
\\


\begin{figure}[h]
\includegraphics[width=9.5cm]{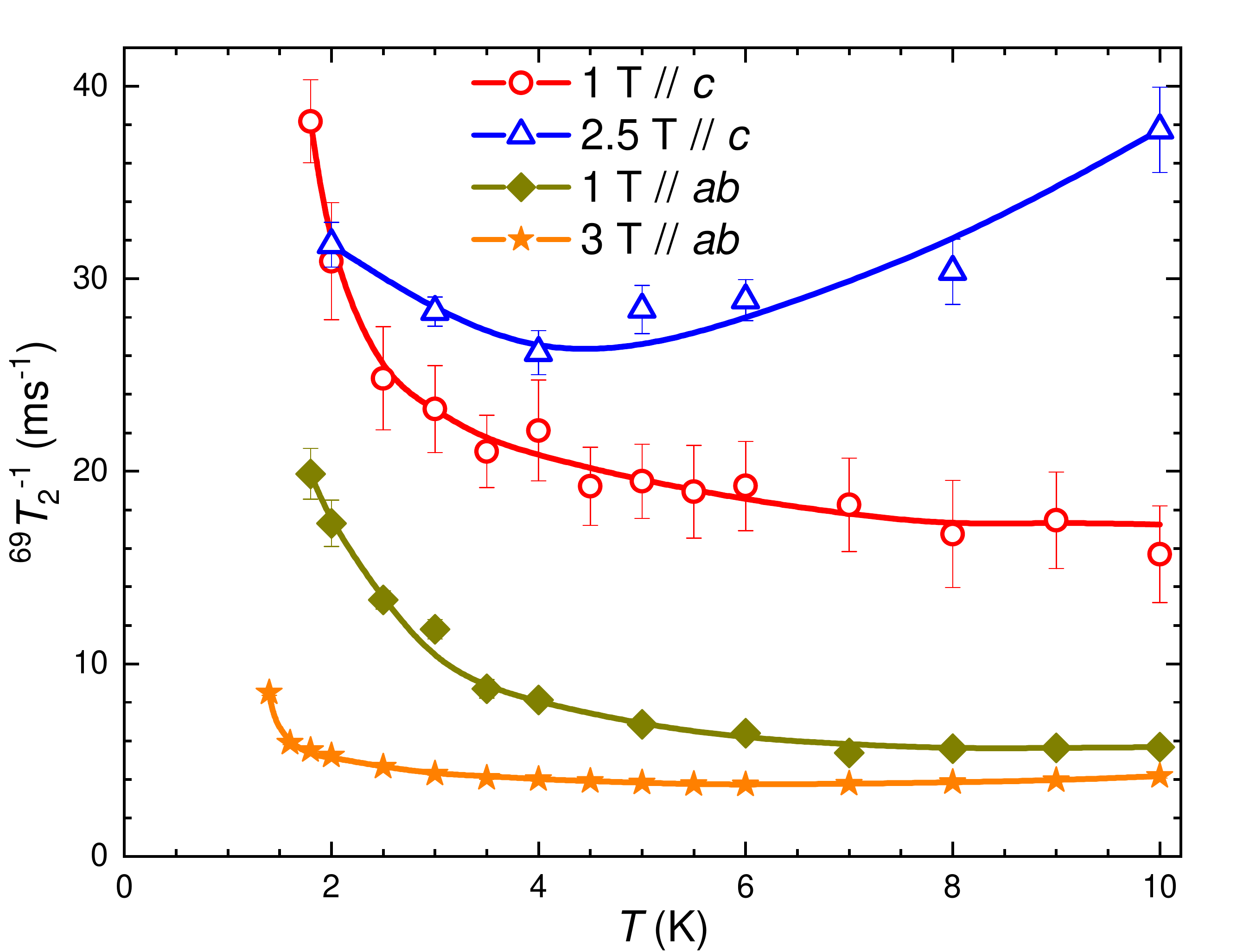}
\renewcommand{\figurename}{\textbf{Supplementary Figure}}
\caption{The spin-spin relaxation rate $1/^{69}T_2$ as functions of
temperatures under various in-plane (//$ab$) and out-of-plane (//$c$) fields.}
\label{fig:ssrr}
\end{figure}

\textbf{Supplementary Note 5: Spin-spin relaxation
rate $1/^{69}T_2$}

The spin-spin relaxation rate $1/^{69}T_2$ with two
field orientations are demonstrated in Supplementary Fig.~\ref{fig:ssrr}.
For the in-plane field $H$//$ab$,
the $1/^{69}T_2$ data increase rapidly
upon cooling below 3~K, which is consistent with the
behaviours of $1/^{69}T_1$,
indicating the onset of dominent contributions from
the magnetic correlations.
As the field increases from 1~T to 3~T,
the $1/^{69}T_2$ data decrease
(meaning longer spin-spin relaxation time).
On the contrary, for the out-of-plane fields
$H$//$c$, the $1/^{69}T_2$ data (above 2~K)
increase when the field changes from 1~T to 2.5~T.
These values are larger than those for the in-plane fields.
In general, the spin-spin relaxation rate $1/T_2$ determines the
decay rate of the NMR signal, and thus the spectra measurements
become very challenging when $1/T_2$ is very large.
\\


\begin{figure}[h]
\includegraphics[width=9.5cm]{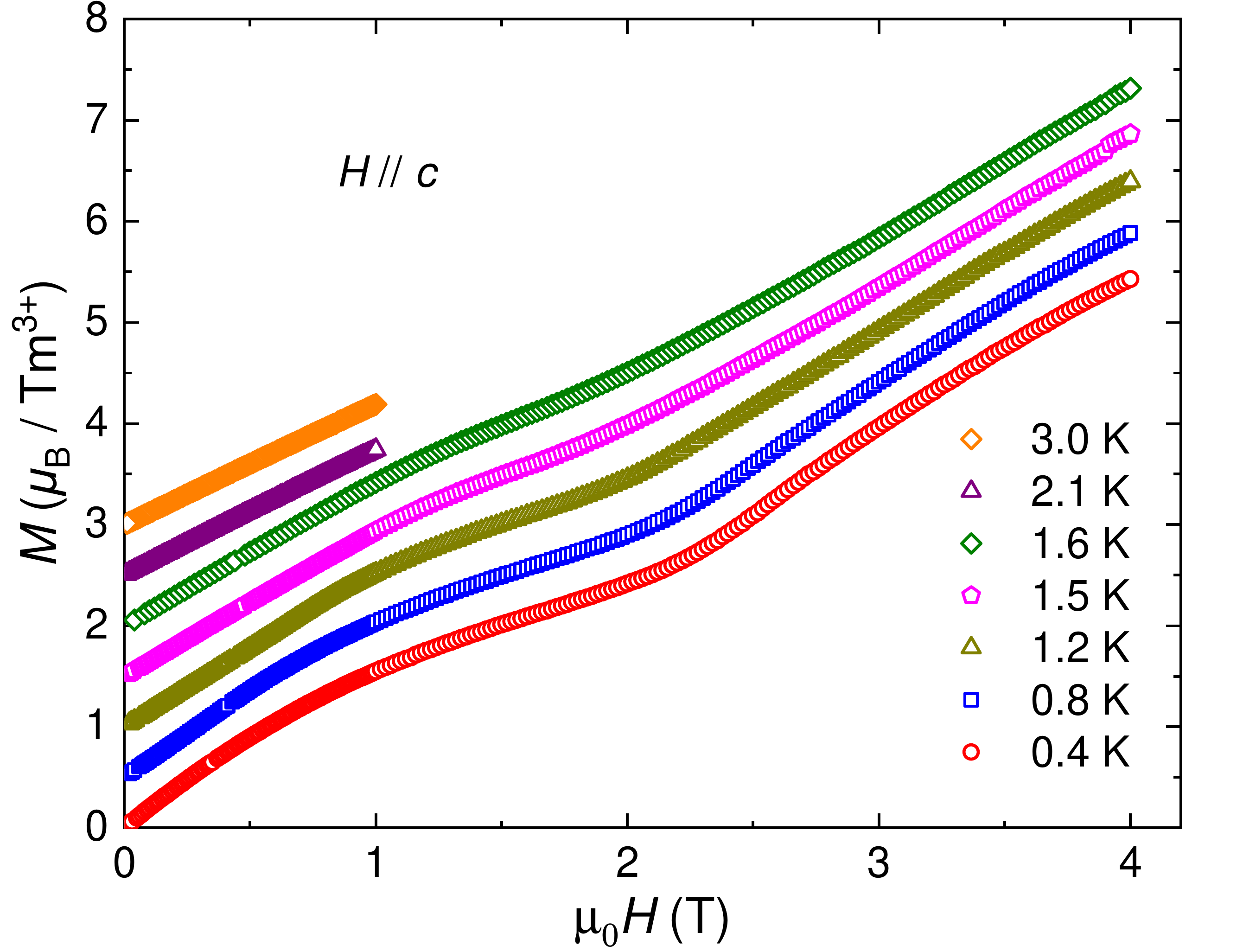}
\renewcommand{\figurename}{\textbf{Supplementary Figure}}
\caption{The magnetization as functions of fields
at selected temperatures. The measured data at
different temperatures are shifted vertically for clarity.}
\label{fig:mh}
\end{figure}

\textbf{Supplementary Note 6: Magnetization curves at
selected temperatures}

The magnetization $M$ of the sample was measured as
functions of fields at various temperatures, as shown
in Supplementary Fig.~\ref{fig:mh}. The magnetization $M$ increases
with fields for all temperatures. At the lowest temperature 0.4~K,
there are two inflection points at about $\mu_0H=$~1~T and
$\mu_0H=$~2.5~T, respectively. The corresponding differential
susceptibility $dM/dH$ are shown and discussed in the
main text.
\\


\begin{figure}[h]
\includegraphics[width=9.5cm]{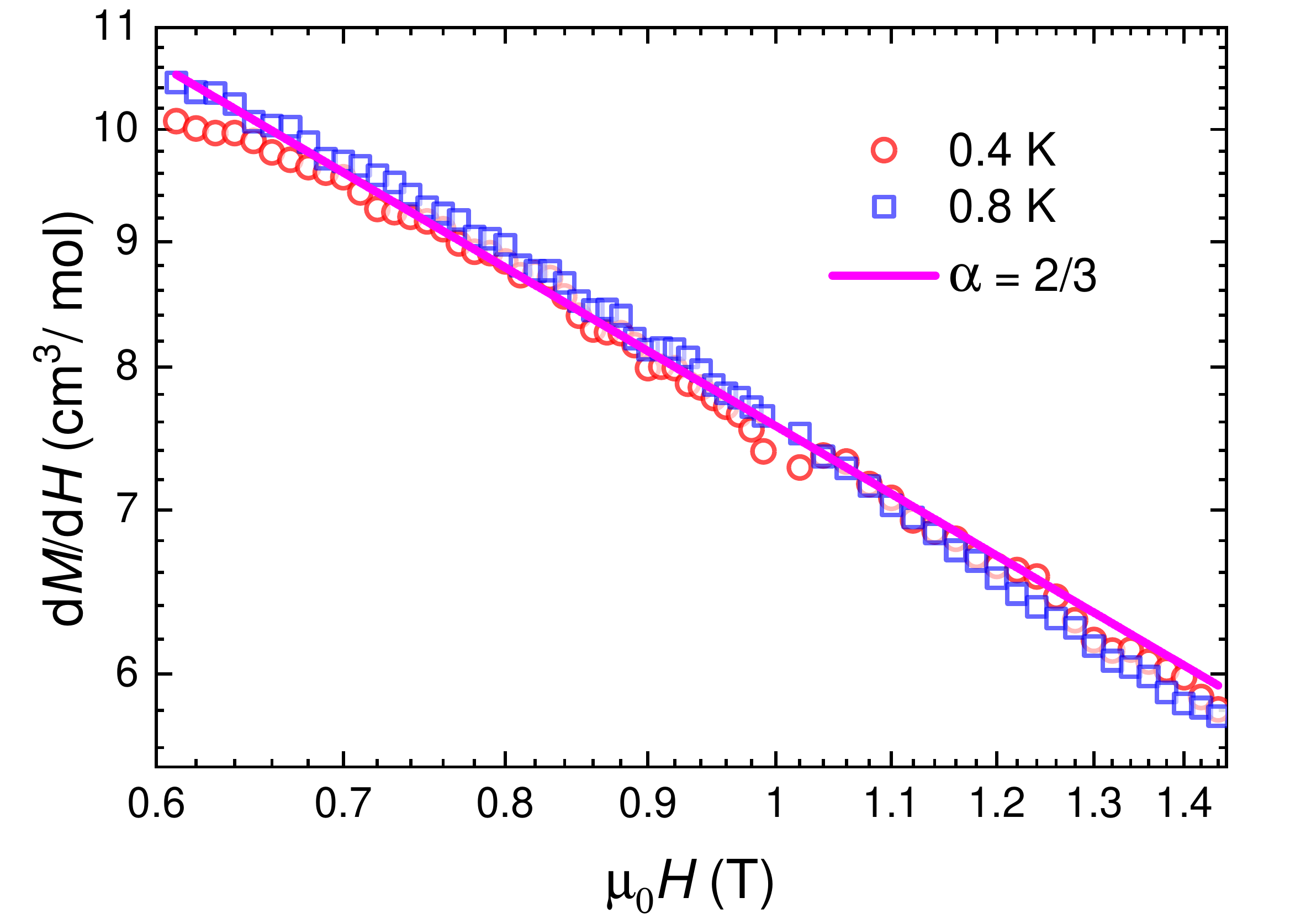}
\renewcommand{\figurename}{\textbf{Supplementary Figure}}
\caption{Fittings of differential susceptibility
at a larger field range. The solid line is the power-law function $dM/dH\sim T^{-2/3}$.}
\label{fig:fitsus}
\end{figure}

\textbf{Supplementary Note 7: Differential susceptibility scalings}

In both experimental and simulated data,
we find only a limited parameter window to observe
the power-law scaling of differential susceptibilities.
In Supplementary Fig.~\ref{fig:fitsus},
we show that the scaling actually extends to
a wider field range [0.6~T, 1.4~T], where we find
the scaling works equally well. The fitted
exponent $\alpha=2/3$, in agreement with
Fig.~3c in the main text, indeed satisfies the 
field-theoretical expectation and constitutes a strong evidence
for the existence of BKT transition in the material
TmMgGaO$_4$.

\end{document}